\documentclass[a4paper,11pt]{article}
\usepackage{bm}
\usepackage{float}
\usepackage{latexsym}
\usepackage{dcolumn}
\usepackage{amsfonts,amssymb}
\usepackage{graphicx}
\usepackage{epsfig}
\usepackage{psfrag}
\usepackage{nccmath}
\usepackage{moresize}
\usepackage{enumerate}
\usepackage[hang,flushmargin]{footmisc}
\usepackage[titletoc,toc]{appendix}
\usepackage{lipsum}
\usepackage{subcaption}
\usepackage{hyperref}
\usepackage{rotating}
\usepackage{multirow}
\usepackage{multicol}
\usepackage{xfrac}
\usepackage{mathtools}
\usepackage{nccmath}
\usepackage{cite}
\usepackage{placeins}
\usepackage{tikz}
\usetikzlibrary{positioning}

\graphicspath{{figures/}} % Location of the graphics fileshttps://www.overleaf.com/project/608a9528e951a97ae206ea82
\usepackage[font=small,labelfont=bf]{caption}
\usepackage{wrapfig}
\usepackage{authblk}
\usepackage{cite}
\usepackage[utf8]{inputenc}
\usepackage[T1]{fontenc}
\usepackage{enumitem}
\usepackage{blindtext}

\title{ \bf Observational constraints on the massive neutrinos induced late-time cosmic acceleration}

\author[1]{Mohit K. Sharma\thanks{mr.mohit254@gmail.com}}
\affil[1]{\small \it Department of Physics \& Astrophysics, University of Delhi, Delhi-110007, India}
\author[2]{Shibesh Kumar Jas Pacif\thanks{shibesh.math@gmail.com}}
\affil[2]{\small \it  Centre for Cosmology and Science Popularization(CCSP), SGT University, Delhi-NCR,
Gurugram 122505, India}
\author[3,4]{Shynaray Myrzakul \thanks{srmyrzakul@gmail.com}}
\affil[3]{\small \it  Center for Theoretical Physics, Eurasian National University, Astana 010008, Kazakhstan}
\affil[4]{\small \it Ratbay Myrzakulov Eurasian International Centre for Theoretical Physics, Nur-Sultan 010009, Kazakhstan. }
\author[3,4]{ Zamzagul Shanina}

\begin{document}
\date{}
\maketitle
\begin{abstract}
 We study a scenario based upon a mass-less $\lambda \phi^4$ theory 
 coupled to massive neutrino matter with $Z_2$ symmetry using a conformal coupling, 
 $A(\phi)=1-\alpha\phi^2/2M_{pl}^2;~\alpha=M^2_{pl}/M^2$ where $M$ is a cut off mass. The 
 chosen coupling generically leads to the spontaneous symmetry breaking at late times such 
 that the field acquires non-zero mass, $m_\phi=(\alpha \Omega_{0\nu})^{1/2}H_0 \ll H_0 $
 and rolls slowly around the true ground state which emerges after spontaneous symmetry 
 breaking. For the statistical analysis, we utilize Pantheon+Multi-Cycle Treasury and OHD 
 data sets. We find that even a small fraction of the neutrino matter density together with 
 its coupling to the scalar field can actually make our model to behave like a weakly 
 dynamical dark energy and have $\Lambda$CDM model as a limiting case.
\end{abstract}

\section{Introduction}

Symmetry breaking is generic to our universe which has gone through various phase transitions starting from GUT era. During each phase transition, part of the symmetry is lost and a rearrangement of the ground state of the system takes place. Being inspired by the success of this paradigm in the early universe, efforts were made to obtain late time accelerate due to spontaneous symmetry breaking in a scenario known as ``Symmetron''. The model is based upon $\lambda \phi^4$ theory with wrong mass sign directly coupled to matter\cite{Amend,Hinterbichler:2010es,Pietroni:2005pv,Olive:2007aj,Bamba:2012yf}. In this framework, the coupling is directly proportional to the trace of the energy momentum tensor of matter. This type of coupling may be induced by a conformal transformation from Jordan to Einstein frame.
The underlying symmetry in the symmetron model is $Z_2$ symmetry which is exact
at early times when matter density is large but breaks down at late time when matter density becomes comparable to the critical  density. The introduction of coupling gives rise to a minimum in the field potential after symmetry breaking where the field could settle  down giving rise to late time acceleration\cite{Riess:1998cb,Perlmutter:1998np,Sahni,Copeland,paddy,BambaDE}.

However, any direct coupling of field to matter is subject to local gravity constraints which are stringent and force the mass of the scalar field to be much larger than the Hubble constant $H_0$ in the true ground state that emerges after breaking  of of $Z_2$ symmetry . Obviously, in this case, the field does not support roll slowly around the minimum, it rather keeps overshooting it  {\it {\`{a} la}}  a ``no-go'' to late acceleration in this framework.

It was pointed out in Ref.\cite{shn}, see Review.\cite{SSrev} for details (also see Refs.\cite{others,Amendola,Wet,H1} on the related theme) that the problem can be avoided by assuming the scalar field coupling to massive neutrino matter. Interestingly, in this case, coupling is absent at early times as massive neutrinos are relativistic there with vanishing trace of their energy momentum tensor. Coupling builds up dynamically only at late stages giving rise to breaking of $Z_2$ symmetry. Since the order of magnitude of neutrino masses is not far from the the characteristic scale of dark energy, the observed late time acceleration can be achieved with little tuning of a free parameter present in the conformal coupling.
Indeed, in this framework, the mass of the field in the ground state is proportional to $\Omega_{0\nu}^{1/2}H_0$.
It was, therefore, argued in Ref.\cite{shn}, that the scalar field rolls slowly around the true ground state that emerges after spontaneous symmetry breaking and might account for the late time acceleration. 
In this paper we subject our model parameters to observational constraints using  OHD \cite{Hubble} and Pantheon+Multi-Cycle Treasury (MCT) \cite{CCH} data sets to confirm the assertion. 

The paper is organized as follows: ($1$) we give brief description of the coupled equations of motion in the flat Friedmann-Robertson-Walker (FRW) spacetime. ($2$) Then, we show that in the effective picture the potential of the scalar field gets modified at the expense of eliminating the contact coupling between neutrino and scalar field. ($3$) For a power-law solution of the scalar field, we derive the background equations in an analytical form. ($4$) At last we carry out our estimations using Supernovae (SN) 1a constraints on the dimensionless Hubble parameter and the standard OHD data sets from various surveys. 
 
\section{The mass-less $\lambda \phi^4 $ directly coupled to massive neutrino matter and FRW evolution equations}

Let us begin with a neutrino-scalar field interacting action in presence of a dust-like matter. 
\begin{equation} \label{action}
\mathcal{S} = \int d^4x \sqrt{-g} \left[ \frac{{M_{pl}^2}}{2}R -
 \frac{1}{2} \partial _\mu \phi \partial ^\mu \phi - V(\phi) \right] +\mathcal{S}_m +\mathcal{S}_\nu(A^2(\phi) g_{\mu\nu}, \Psi_\nu) \, ,
\end{equation}
where $g$ is the determinant of four-dimensional metric $g_{\mu\nu}$, $R$ is the Ricci scalar, $\phi$ is the scalar field, $V(\phi)$ is the scalar field
potential and $\mathcal{S}_m$ and $\mathcal{S}_\nu$ are actions for the
matter and neutrino, respectively. The requirement for the spontaneous symmetry breaking for the mass-less scalar field and to comply with $Z_2$ symmetry, we make choice of $V(\phi)$ and $A(\phi)$ as follows:
\begin{equation} \label{potential}
    V(\phi) = \frac{\lambda }{4}\phi^4 \, , \quad A(\phi) = 1-\frac{\alpha \phi^2}{2 M_{pl}^2} \, ,
\end{equation}
where $\alpha$ is some constant. In fact, $\alpha \equiv M_{pl}^2 / M^2$ fixes the cutoff scale and can be constraint using the observational data. In the FRW spacetime : $ds^2 = -dt^2 + \delta_{ij}dx^i dx^j$, the Friedmann equation can be expressed as
\begin{equation} \label{Friedmann}
    H^2 = \frac{1}{3  M_{pl}^2} \left[\rho_m + \rho _\nu +\frac{1}{2} \dot{\phi}^2 +V(\phi)\right] \, ,
\end{equation}
where $\rho_m$ and $\rho_\nu$ are matter and neutrino energy densities, respectively, and the overdot denotes the derivative with respect to cosmic time $t$. One derives the equation of motion for scalar field as
\begin{equation} \label{field_equation}
\ddot{\phi} +3 H \dot{\phi} = -V_{,\phi} + \frac{A_{,\phi}}{A} T_\nu
\end{equation}
where $T_\nu(=\!-\rho_\nu +3p_\nu)$ ($p_\nu$: pressure of neutrino) is the trace of the neutrino matter sector. Due to the contact coupling between neutrino and scalar field, the individual energy densities of both do not remain self-conserved anymore, however the overall energy density for all components is conserved. 
\begin{equation} \label{cons-eqn}
    \dot{\rho}_\nu + 3 H(\rho_\nu + p_\nu)= \frac{A_{,\phi}}{A} \dot{\phi}
    (\rho_\nu - 3 p_\nu) \, .
\end{equation}
As neutrino gets decoupled from other species at very early-times (when the temperature of the universe is around $1$MeV), since then they are expanding freely in the universe. Also as their masses are nearly ranges from $\mathcal{O}(10^{-1}-10^{-2})$, they behaves like a non-relativistic dust at late-times. Since, our objective is to study the late-time behavior, therefore, we take $p_\nu=0$ throughout our analysis. As a consequence, Eq. (\ref{field_equation}) can be re-written as
\begin{equation} \label{field-eqn2}
    \ddot{\phi}+3H\dot{\phi} =  -V_{,\phi} -A_{,\phi} \hat{\rho}_\nu
\end{equation}
where $\hat{\rho}_\nu= a\rho _\nu$ is an `effective' neutrino matter density which minimally coupled with the scalar field and is conserved : 
\begin{equation} \label{rho_nu}
\dot{\hat{\rho}}_\nu+3H\hat{\rho}_\nu=0   \, .
\end{equation}
With this redefinition, one gets a modified scalar field potential. In particular, the effect of the interaction now gets appeared in the effective potential $V_{\text{eff}}(\phi)$. It is this modification, which makes the scalar field potential to give-rise to late-time cosmic acceleration and field behaving as a dark energy (DE). Hence, the Eq. (\ref{field-eqn2}) can be re-stated as
\begin{equation} \label{effpotential}
 \ddot{\phi}+3H\dot{\phi} = - V_{\text{eff}}(\phi) \, , \quad \mbox{where} \quad
    V_{\text{eff}}(\phi) = \frac{\lambda}{4}\phi^4 - \frac{\alpha \hat{\rho}_\nu}{2 M_{pl}^2}\phi ^2 \, .
\end{equation}
In the effective framework, the Friedmann equation can be written in terms of effective neutrino energy density $\hat{\rho}_\nu$  and field energy density $\rho_\text{eff}(\phi)$ as
\begin{equation} \label{Friedmann2}
    H^2(a)=\frac{1}{3 M_{pl}^2} \left[\frac{\rho_{0m}}{a^3} + \frac{\hat{\rho}_{0\nu}}{a^3} + \rho_\text{eff}(\phi) \right] \, \quad \mbox{where,} \quad \rho_\text{eff}(\phi)=\frac{\dot{\phi}^2}{2} 
    + \frac{\lambda}{4}\phi^4 - \frac{\alpha \hat{\rho}_\nu}{2 M_{pl}^2}\phi ^2 \, .
\end{equation}
Let us define the dimensionless density parameter for matter, neutrino and scalar field as
\begin{equation} \label{density-params}
\Omega_m \equiv \frac{\rho_m}{3H^2M_{pl}^2} \, , \quad
\Omega_\nu \equiv \frac{\hat{\rho}_{\nu}}{3H^2M_{pl}^2} \, , \quad 
\mbox{and} \quad \Omega_\phi \equiv \frac{\rho_\text{eff}(\phi)}{3 H^2 M_{pl}^2} \, .
\end{equation}
In order to solve the set of equations (\ref{rho_nu}), (\ref{effpotential}) and (\ref{Friedmann2}), let us assume a solution for the scalar field in terms of a simple power-law form:
\begin{equation} \label{ansatz}
    \phi(t) = \phi_0 a^n(t)\, , \quad \mbox{where} \quad \phi_0 = \phi|_{a=1} \, ,
\end{equation}
and $n$ is a constant. Using Eqs. (\ref{Friedmann2}), (\ref{density-params}) and (\ref{ansatz}), one can extract out 
self-coupling parameter of scalar field i.e. $\lambda$ from the present value of field density parameter: $\Omega_{0\phi}=\Omega_{\phi}|_{a=1}$ as 
\begin{equation} \label{Omega-phi}
\Omega_{0\phi} =\frac{\rho_\text{eff}(\phi)|_0}{3H_0^2 M_{pl}^2} \, \quad 
\mbox{implies} \quad \lambda = \frac{2H_0^2(3\alpha \phi_0^2\Omega_{0\nu}+6M_{pl}^2\Omega_{0\phi})}{\phi_0 ^4} \, .
\end{equation}
By defining $\phi_0 := \sigma M_{pl}$ (where $\sigma$ is a dimensionless field value at the present epoch) one can re-write the Friedmann equation as
\begin{equation}
H\!=\!H_0\left( \frac{6(\Omega_{0\phi}\!-\!1) -3 \alpha\sigma^2Ba^{2n} +[(\sigma^2(n^2 +
3\alpha B)-6\Omega_{0\phi}]a^{3+4n})}{(n^2 \sigma^2 a^{2n}-6)a^3} \right)^{1/2}
\end{equation}
where 
$\Omega_{0m}=\Omega_m|_{a=1}$, $\Omega_{0\nu}=\Omega_\nu|_{a=1}$ and $B:=\Omega_{0\phi}+\Omega_{0m}-1$. From the constraint relation: $\Omega_{0\phi}+\Omega_{0m}+\Omega_{0\nu}=1$ one can note that $B$ is equivalent to the neutrino density parameter which we have eliminated by using the constraint for our parametric estimation.

Similarly, the effective DE equation of state $w_\text{DE}$ can be worked out as
\begin{eqnarray} \label{eqnofstate}
& w_\text{DE}(a)  := \frac{p_\text{eff}(\phi)}{\rho_\text{eff}(\phi)} =\frac{\dot{\phi}^2-2V_{\text{eff}}(\phi)}{\dot{\phi}^2+2V_{\text{eff}}(\phi)} \\&= 
\frac{-H_0^2\left( \sigma^2(n^2+3\alpha B\right)a^{3+2n}+\sigma^2\left(3H_0^2\alpha B-n^2a^3 H(a)^2\right)}{H_0^2\left( \sigma^2(n^2+3\alpha B\right)a^{3+2n}+\sigma^2\left(3H_0^2\alpha B-n^2a^3 H(a)^2\right)}
\, .
\end{eqnarray}
Note that in the limit $n\to 0$, when field becomes constant, 
one gets back the $\Lambda$CDM model with $w_\text{DE}=-1$. 
However, at present when $a(t)=1$, the effective DE equation of state is given as
\begin{equation} \label{wde0}
w_\text{DE}^{(0)} = -1 + \frac{n^2 \sigma^2}{3\Omega_{0\phi}} \geq -1 \, .
\end{equation}
The mass of the field $m_\phi(\equiv d^2 V_{\text{eff}}(\phi)/d \phi^2)$ in the true ground state is expressed through $\alpha$ and $\Omega^0_\nu$ as
\begin{equation} \label{massphi}
m^2_\phi= 6\alpha \Omega_{0\nu} H^2_0  
\end{equation}
which should be much smaller than $H_0$ in order to support
the slow roll required by the observed late time cosmic acceleration. In the next section, we will carry out parametric estimations for the setup.

\section{Parametric estimations with Pantheon+MCT and $H(z)$ datasets}

\begin{figure}[htp]

\centering
\includegraphics[width=.46\textwidth]{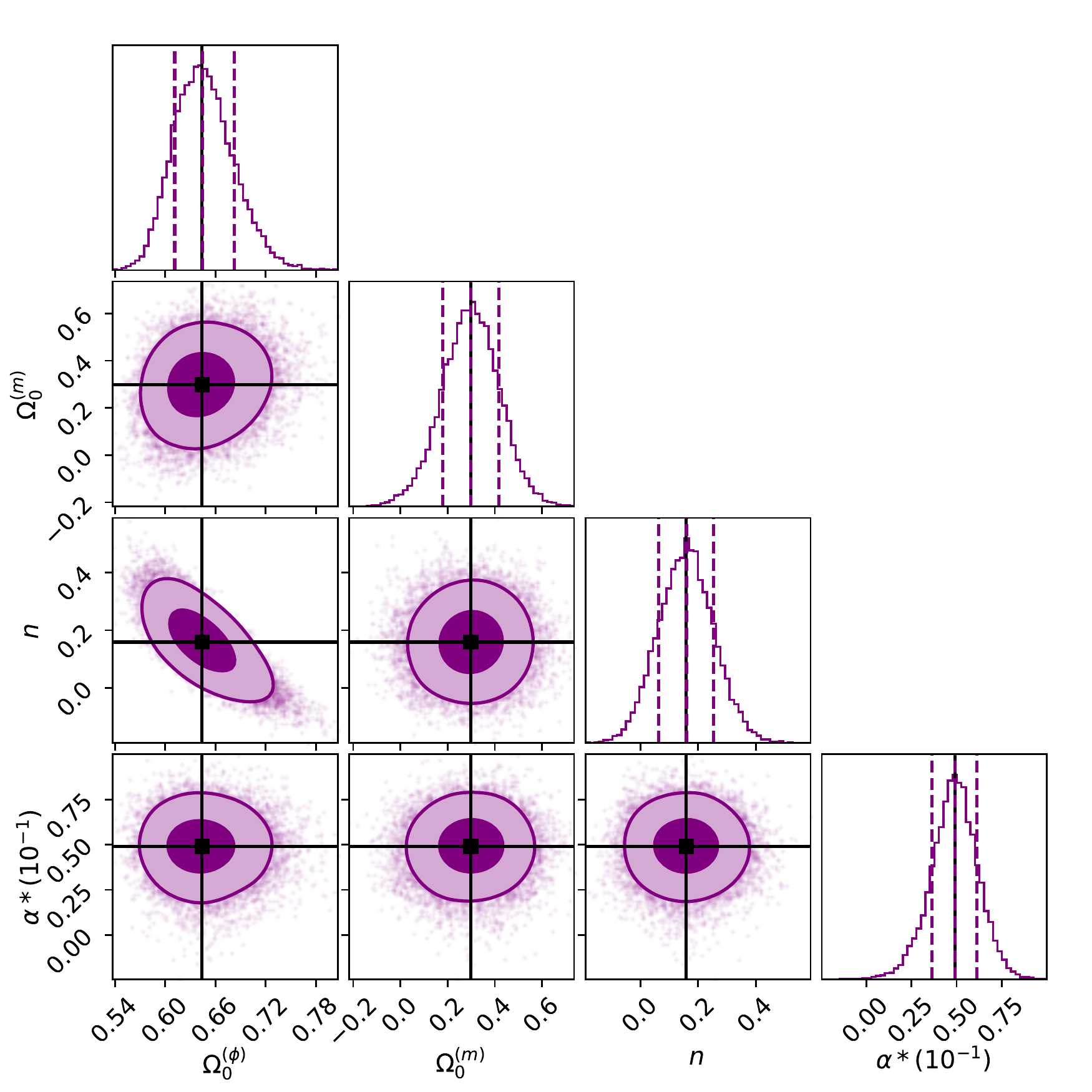} \hfill
\includegraphics[width=.46\textwidth]{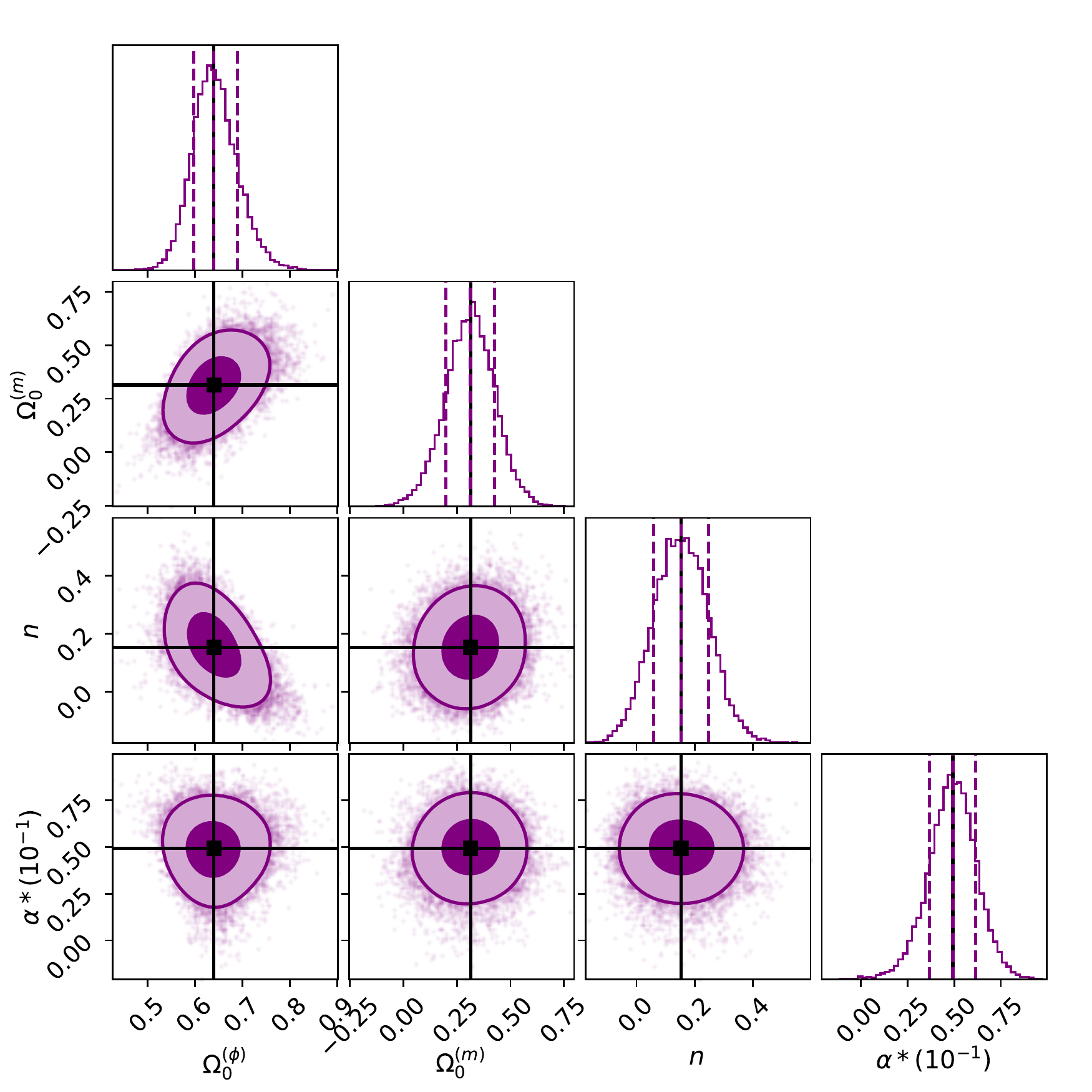} \hfill
\includegraphics[width=.46\textwidth]{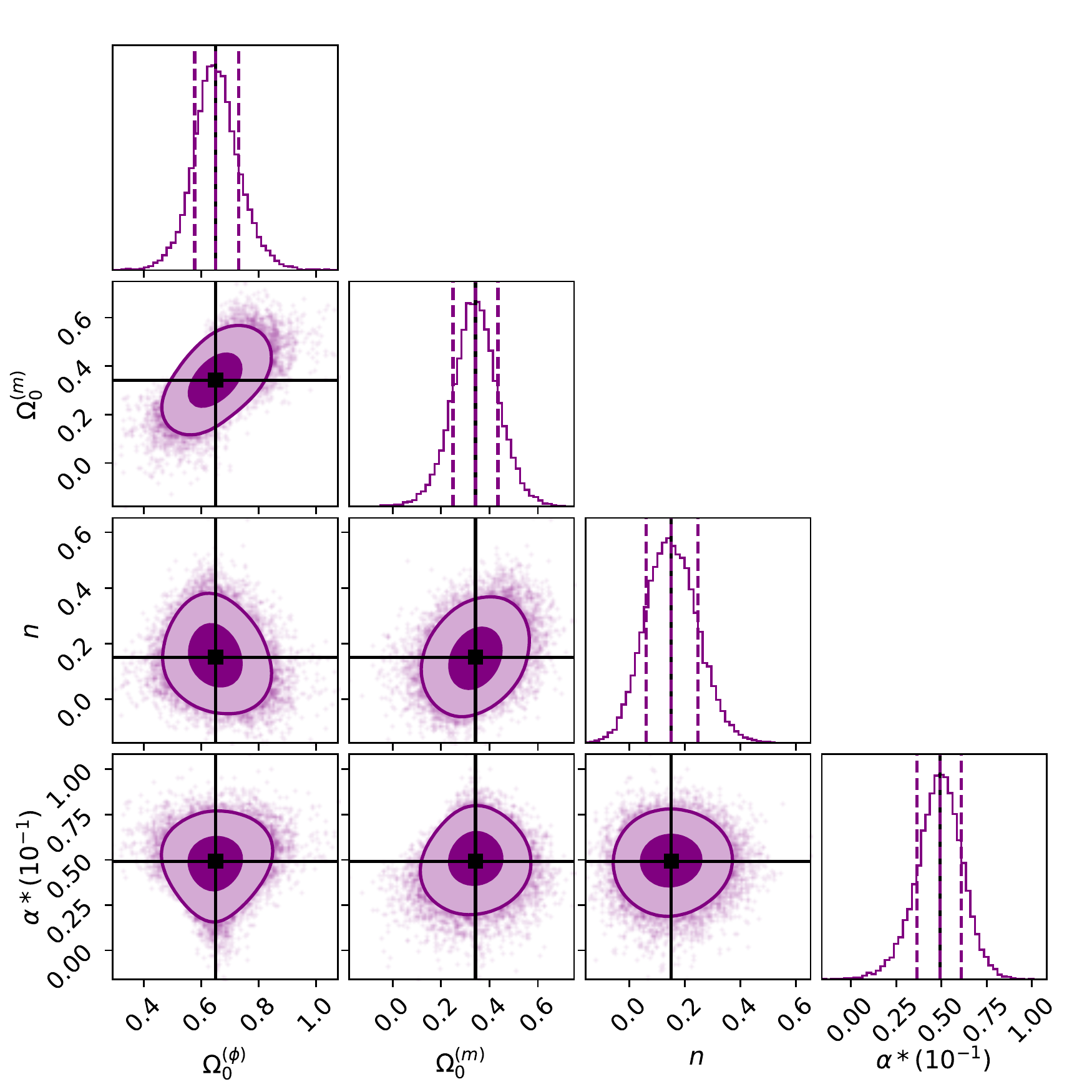} \hfill
\includegraphics[width=.46\textwidth]{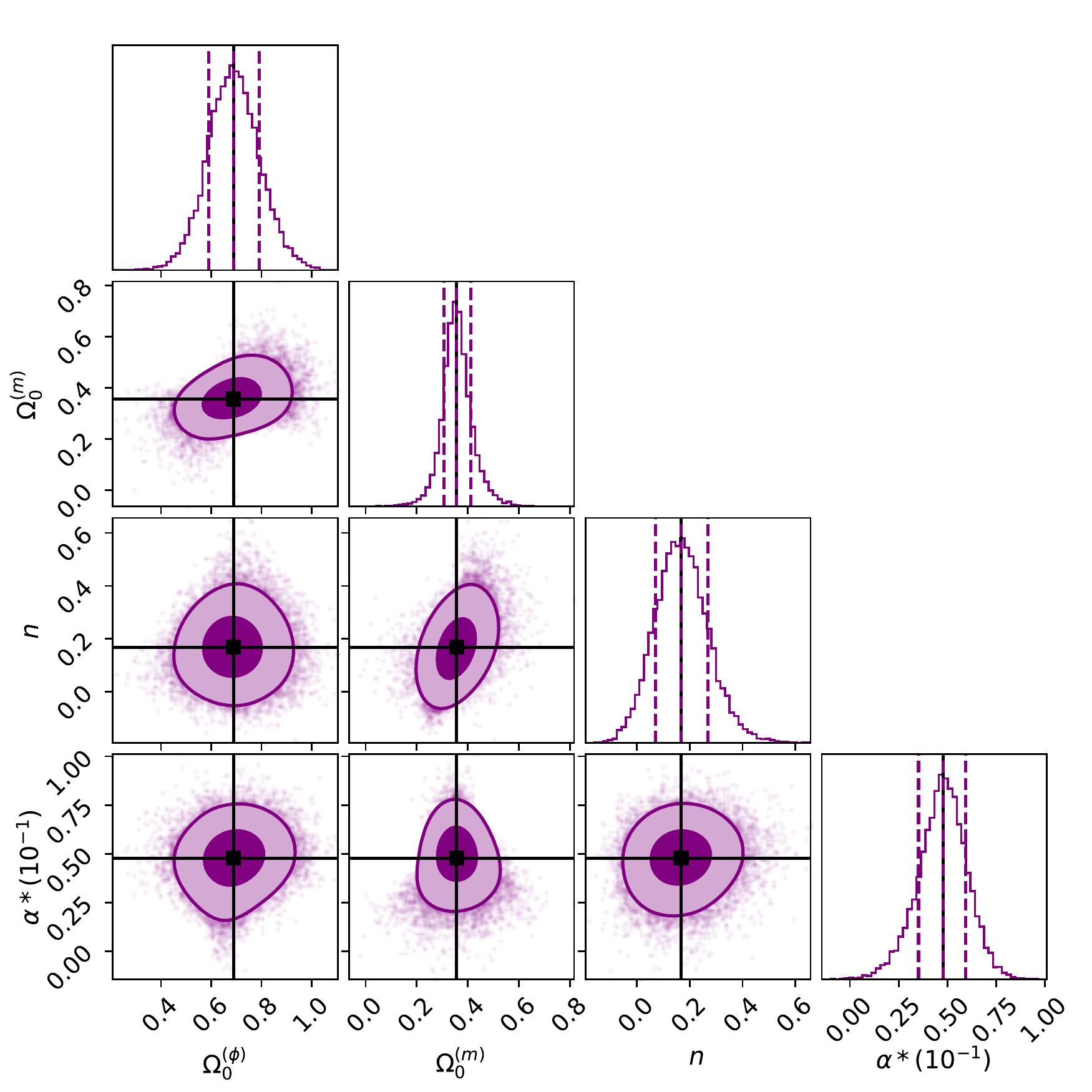} \hfill
\caption{$2-D$ posterior distribution for Pantheon+MCT and $H(z)$ dataset upto $2\sigma$ limit for different values of $\sigma$. The upper-left, upper-right, lower-left and lower-right figures corresponds to $\sigma=0.5,1,1.5$ and $2$, respectively. The black line represents the best fit value and the dotted-line represents the corresponding $1\sigma$ confidence interval.}
\label{fig:contour}

\end{figure}

\begin{figure}[htp]

\centering
\includegraphics[width=.43\textwidth]{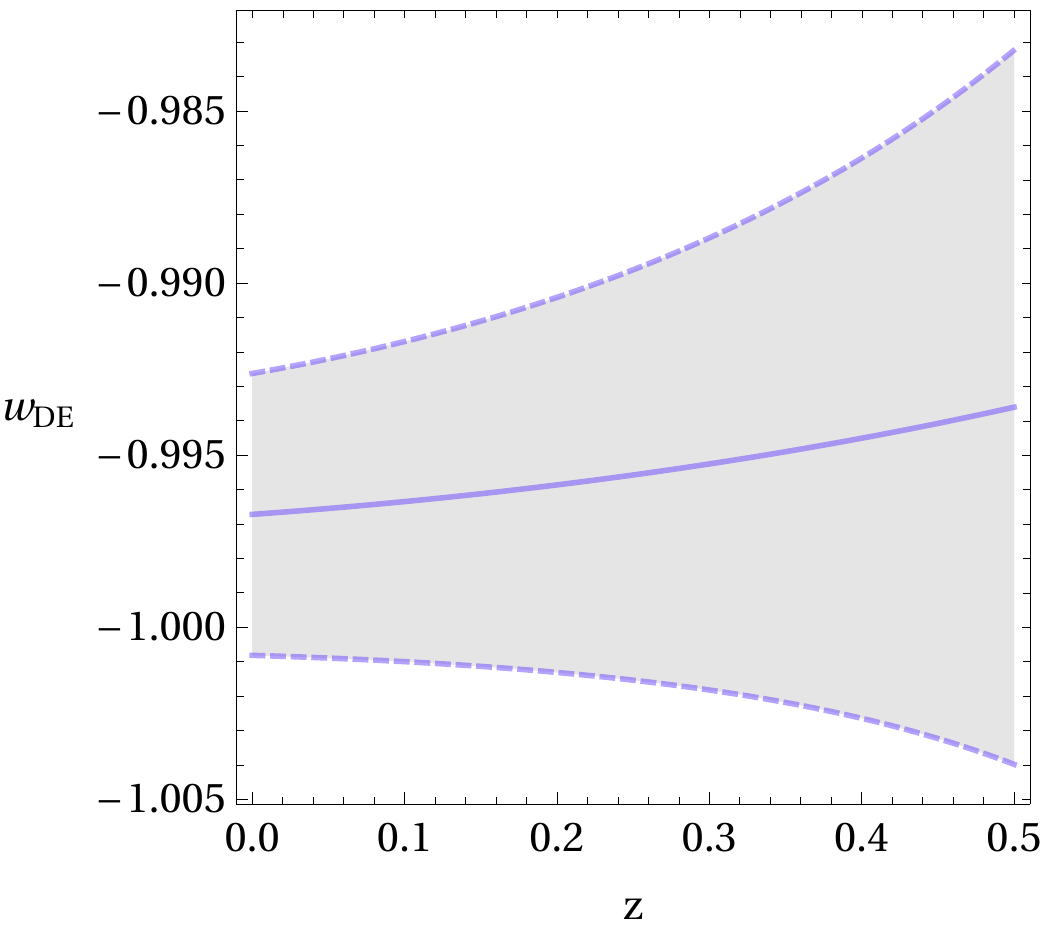} \hfill
\includegraphics[width=.43\textwidth]{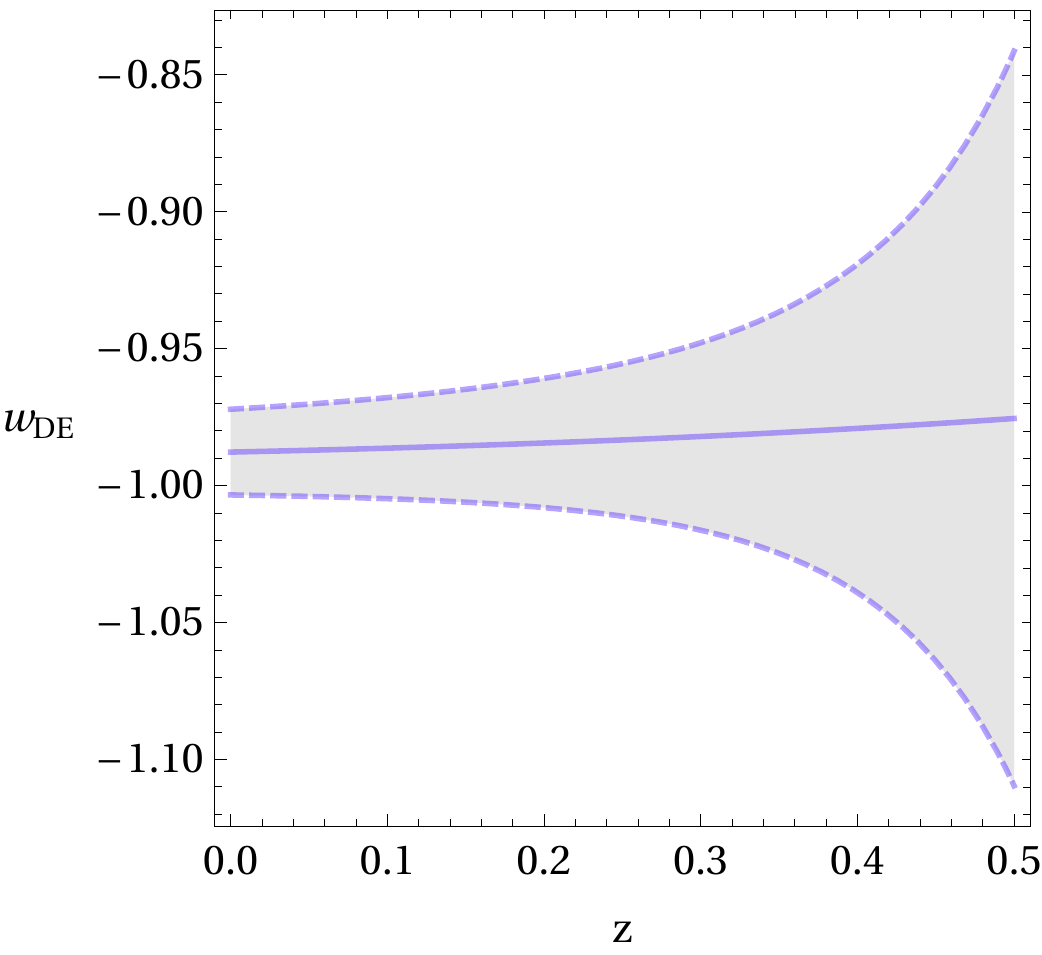} \hfill
\includegraphics[width=.43\textwidth]{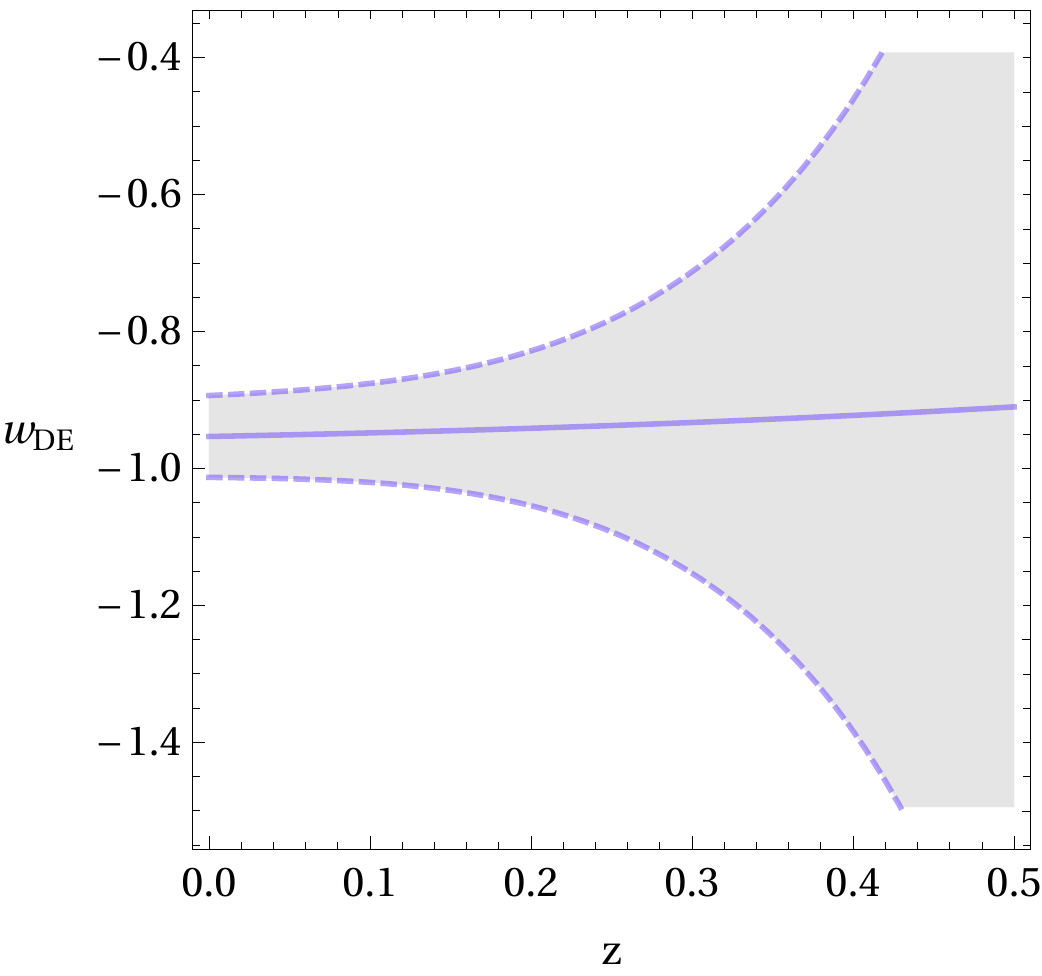} \hfill
\includegraphics[width=.43\textwidth]{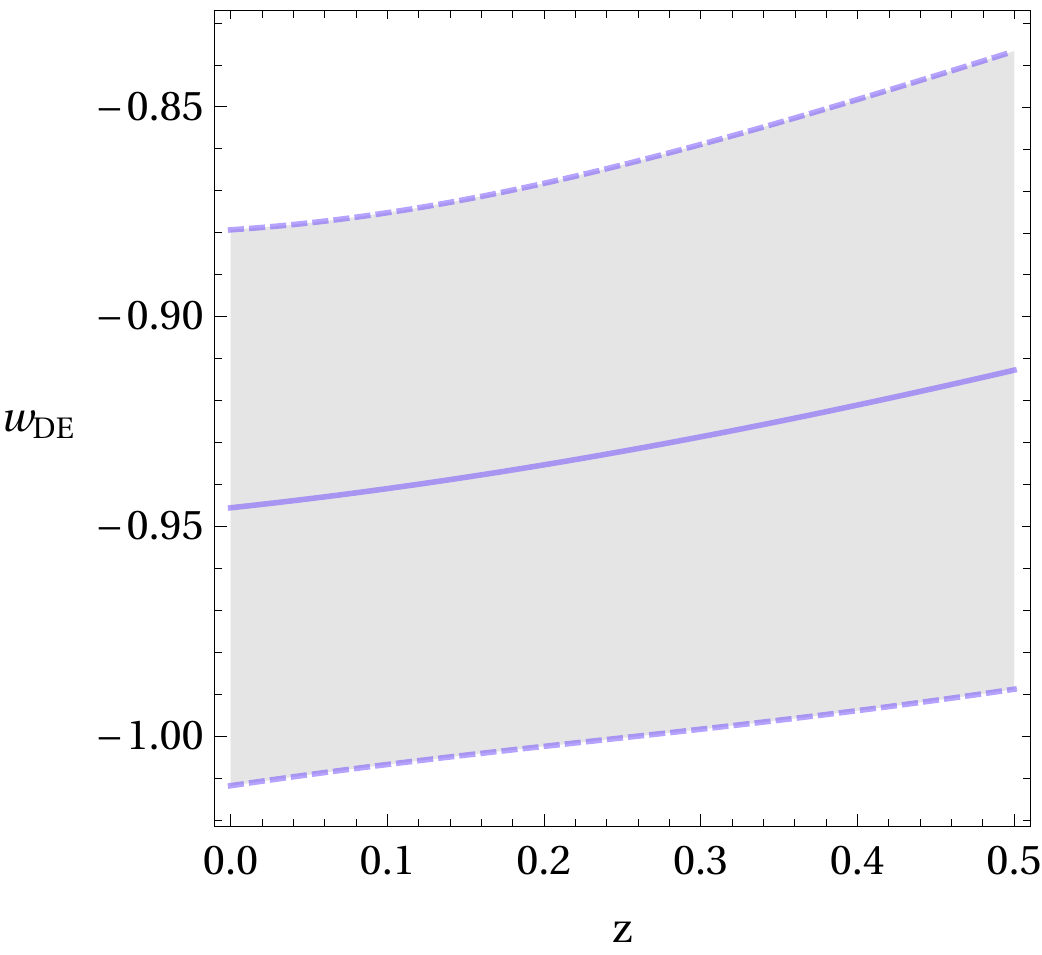} \hfill
\caption{Evolution profile of DE effective equation of state with red-shifts $z\in[0,1]$. The upper-left, upper-right, lower-left and lower-right figures corresponds to $\sigma=0.5,1,1.5$ and $2$, respectively. The solid (blue) line represents the best-fit and dashed-lines represent the $1\sigma$ level.}
\label{fig:wxplot}
\end{figure}

\begin{figure}
    \centering
    \includegraphics[width=.55\textwidth]{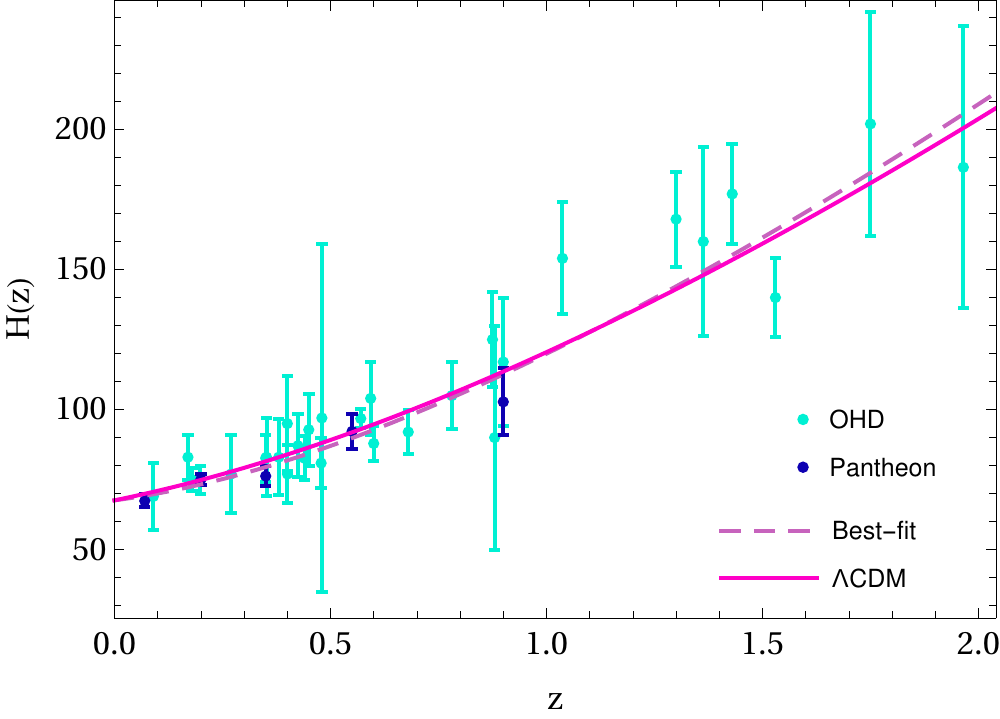}
    \caption{Reconstructed best-fit profile of $H(z)$ corresponds to $\sigma=1$ and $H_0=67.6$ Km/s/Mpc as well as for the $\Lambda$CDM model
    ranges from $z \in [0,2]$. The OHD error bars are shown in light blue, and the Pantheon+MCT ones are in dark. }
    \label{fig:H}
\end{figure}

For our estimations, we use the combination of OHD and Pantheon+MCT dataset. In particular, we perform the Metropolis-Hastings technique for the Markov Chain Monte Carlo (MCMC) simulation on the combined $\chi^2$, which is defined as
\begin{equation} \label{chisquare}
    \chi^2 := \sum_i (X-X(z_i))\cdot C_{ij}^{-1} \cdot (X-X(z_i))
\end{equation}
where $X$ and $X(z_i)$ are theoretical and observed quantities, respectively,  and $C_{ij}$ is the co-variance matrix between five data points of $E(z):=H(z)/H_0$, given in \cite{CCH}. We fix Hubble constant at the combined Planck+Lensing+BAO best-fit i.e. $H_0=67.6$ km/s/Mpc \cite{Planck18-CP}. Also, we repeat our analysis four different values of the present value of dimensionless field parameter i.e. $\sigma =0.5,1,1.5$ and $2$. The prior ranges given for remaining parameters are:
\begin{equation} \label{priors}
    0.5 \leq \Omega_{0\phi} \leq 0.9 \, \quad 0.1 \leq \Omega_{0m} \leq 0.5 \,  \quad 0 \leq n \leq 0.5 \, \quad 0 \leq \alpha \times 10^{-1} \leq 0.7 \, .
\end{equation}
The obtained parametric dependence between all parameters are shown in Fig. (\ref{fig:contour}) for different values of $\sigma$ and the corresponding best-fit values with their $1\sigma$ levels are shown in table (\ref{E-tab1}). Since the neutrino matter density tends to decrease the effective energy density of the scalar field and hence its density parameter (see Eq.(\ref{Friedmann2})), we, therefore, observe slightly high matter density parameter and small field density parameter. In table (\ref{E-tab1}), one finds that as $\sigma$ increases $\Omega_{0\phi}(\Omega_{0m})$ also increases (decreases). The observational allowed evolution of DE equation of state $w_{DE}$ is shown in fig. (\ref{fig:wxplot}). Here let us emphasize that although for the estimations we show that even within $1\sigma$ limit the DE equation of state can take values less than $-1$, but theoretically it is not possible to have $w_\text{DE}^{(0)}<-1$ (see Eq. (\ref{wde0})). Hence, in order for the estimations to be consistent with the fundamentals of the theory, only the errors which restricts $w_\text{DE}^{(0)}$ $\geq -1$ are feasible. Also for a particular case, i.e. for $\sigma=1$, we also plot the reconstructed best-fit of $H(z)$ in fig. (\ref{fig:H}).

%\begin{figure}[!ht]
%    \centering
%    \includegraphics[scale=0.4]{Neutrino_cont.pdf}
%    \caption{$2-D$ posterior distribution for Pantheon and $H(z)$ dataset upto $2\sigma$ limit. The black line represents the best fit value and the dotted-line represents the corresponding $1\sigma$ confidence interval.}
%    \label{fig:contour}
%\end{figure}
%

\begin{table*}[ht]
\centering
\renewcommand{\arraystretch}{1.8}
{
%\begin{tabular}{||p{3cm}||p{2.3cm}|p{2.3cm}|p{2.3cm}|p{2.3cm}||p{1.5cm}||}
\begin{tabular}{||c|c|c|c|c||c||}
\hline
 & \multicolumn{4}{c||}{\small Pantheon+MCT $+H(z)$ data-set}
& \\
 &  \multicolumn{4}{c||}{\footnotesize (best fit \& $1\sigma$ limits)}  & ~ $\chi ^2_{bf} $ \\
\cline{2-5}{}
%\cline{7-8}
%
 & $\Omega_{0\phi}$ & $\Omega_{0m}$  & $n$ 
 &   $\alpha\times(10^{-1})$ &  \\
%\
\hline\hline
1. $\sigma=0.5$ & $ 0.644^{+0.038}_{-0.033}$ & $0.298^{+0.118}_{-0.120}$  & $0.159^{+0.095}_{-0.096}$ &
 $0.490^{+0.123}_{-0.128} $ & $25.437$   \\
\hline
2. $\sigma=1$ & $ 0.640^{+0.050}_{-0.043}$ & $0.314^{+0.111}_{-0.115}$  & $0.153^{+0.095}_{-0.094}$ &
 $0.493^{+0.122}_{-0.126} $ & $25.484$   \\
\hline
3. $\sigma=1.5$ & $ 0.650^{+0.081}_{-0.073}$ & $0.342^{+0.093}_{-0.093}$  & $0.151^{+0.096}_{-0.090}$ &
 $0.492^{+0.119}_{-0.127} $ & $25.396$   \\
\hline
4. $\sigma=2$ & $ 0.689^{+0.103}_{-0.098}$ & $0.356^{+0.057}_{-0.049}$  & $0.168^{+0.102}_{-0.097}$ &
 $0.478^{+0.116}_{-0.126} $ & $25.477$   \\
%2. {\small GOLD$+ H(z)$ } & $0.2760^{+0.0361}_{-0.0348}$ &  $0.6911^{+0.0216}_{-0.0211}$  & $ 0.0343^{+0.0382}_{-0.0243}$ & 
%$30.7433 $ \\
%
\hline\hline
\end{tabular}
}
\caption{\footnotesize Best fit values of parameters with their $1\sigma$ confidence limits obtained for Pantheon+MCT and $H(z)$ dataset. $\chi^2_{bf}$ represents $\chi^2$ corresponding to the best-fit values of parameters.}
{\label{E-tab1}}
\end{table*} 

%
%From above estimations, one can also constraint $w_\text{DE}^{(0)}$ as,
%
%\begin{equation} \label{wde0-esimate}
%    w_\text{DE}^{(0)} = -0.996 \pm {0.043} \, ,
%\end{equation}
%
Also by Eq. (\ref{massphi}) the dimensionless mass parameter $\hat{m}^2_\phi \equiv m^2_\phi / H_0^2 \simeq  \mathcal{O}(10^{-1})$
which confirms the fact that $m_\phi \ll H_0$ in the ground state that emerges after symmetry breaking and this is consistent with the observed value of $w_\text{DE}^{(0)}$. 
Also note that the combined constraint by Planck and BAO give $\Omega_{0\nu}< 0.003$ at $95\%$ confidence level.
This constraint is consistent with our theoretical arguments that the neutrino density must be very low in order to support the slow roll around the true ground (with little tuning of $\alpha$).
%prevent the large corrections in the effective scalar field potential.  

\section{Conclusion and Discussion}

In this paper, we carry out the observational estimations on the massless $\lambda \phi^4$ scenario which is non-minimally coupled to the neutrino matter (massive) in presence of the pressure-less matter. We have used the conformal coupling, $A(\phi)=1-\alpha\phi^2/2M_{pl}^2;~\alpha=M^2_{pl}/M^2$ such that
at late-times, when neutrinos turn non-relativistic, the neutrino matter starts behaving like a non-relativistic dust whose coupling  to  scalar field builds up dynamically. The contribution of coupling modifies the scalar field potential such that tachyonic instability builds up in the system deriving it from $\phi=0$ to a true ground state where with $\phi\neq 0$ where $m_\phi=(\alpha \Omega_{0\nu})^{1/2}H_0 $. Consequently, not much fine tuning of $\alpha$, in this case, is required to achieve the slow roll required to account for the observed value of the dark energy equation of state parameter (see table (\ref{E-tab1})). 

In order to solve the system of equations we consider a power-law solution for the scalar field, $\phi(t)\propto a(t)^n$. We show that the present effective dark energy equation of state has its minimum at $-1$ (see Eq. (\ref{wde0}), and mimics the cosmological constant $\Lambda$ at late-times.  In this framework, we have two parameters in addition to the cosmological parameters, namely, $n$ and $\alpha$ which control the slow roll evolution of the system, thereby, the observed late time accelerated expansion would constrain both the parameters.

For our estimations, we perform a MCMC technique using the combined dataset of Pantheon+MCT and $H(z)$ on the Friedmann Eq. (\ref{Friedmann2}) for different values of field value $\sigma$ and the parametric dependence within is shown in Fig.(\ref{fig:contour}). The obtained estimations for $n$ and $\alpha\times (10^{-1})$ are shown in table (\ref{E-tab1}), which results the cutoff scale $M$ to be smaller than the Planck Mass $M_{pl}$. One also see that although the neutrino-scalar field coupling slightly enhance the dimensionless matter density parameter by lowering the field density contribution in the universe, the estimate of $\Omega_{0m}$ are still compatible with the latest Planck 2018 results within $1\sigma$ level. Our results demonstrate that with the small value of the neutrino density parameter, the system  exhibits a behavior of a weakly dynamical dark energy with mildly evolving $w_{DE}(z)$ at late times close to the present epoch (acceleration commences when $z=z_{tr}\simeq 0.67$). With the obtained best-fit for $\alpha$ and $\Omega_{0\nu}$, and using Eq.(\ref{massphi}) we have shown that the  mass of the scalar field around the present epoch  is much smaller than $H_0$ expected from the requirement of slow roll which is reflected in the behaviour of dark energy equation of state parameter, see Fig.(\ref{fig:wxplot}). We also depict the observational allowed evolution of DE equation of state with redshift (Fig.(\ref{fig:wxplot}), in which one can see that $w_{DE}(z)$ is indeed mildly evolving. The evolution of $H(z)$ is shown in Fig.(\ref{fig:H}), together with its corresponding one for the $\Lambda$CDM model. Although, in our formulation and estimations, we eliminate $\Omega_{0\nu}$ in terms of $\Omega_{0\phi}$ and $\Omega_{0m}$ by using constraint equation, one finds that $\Omega_{0\nu} \ll \mathcal{O}(10^{-1})$ which is consistent with the Planck+BAO constraint. We, therefore, conclude that the theoretical scenario based upon the coupling of massive neutrino to scalar field, as an underlying  cause for the late-time cosmic acceleration,  corroborates with observational results. In other words, the spontaneous symmetry breaking due massive neutrino matter coupling that gives rise to slow roll of scalar field around the true minimum, is consistent with observations. Last but not least, we should emphasize that the proper estimation of $\Omega_{0\nu}$ can be obtained by analyzing the suppression of the observable Cosmic Microwave Background (CMB) spectra, which we shall try to address in our future work.

\section{Acknowledgements}
We thank M. Sami for useful discussions. MKS thanks the Centre for Cosmology and Science Popularisation (CCSP) for hospitality where the work was initiated. SM is supported by the Ministry of Education and Science of the Republic of
Kazakhstan, Grant No. BR05236322 and Grant N0 AP08052197. The work of MKS is supported by the Council of Scientific and Industrial Research (CSIR), Government of India.
ZS is supported by the Ministry of Education and  
  Science of the Republic of Kazakhstan, Grant  AP08052197.

\end{document}